# Structural distortion and magnetism of BiFeO$_3$ epitaxial thin films: a Raman spectroscopy and neutron diffraction study

H. BEA[1,*], M. BIBES[2], S. PETIT[3], J. KREISEL[4] and A. BARTHELEMY[1]

[1]Unité Mixte de Physique CNRS-Thales, Route Départementale 128, 91767 Palaiseau, France

[2]Institut d'Electronique Fondamentale, CNRS, Université Paris-Sud, 91405 Orsay, France

[3]Laboratoire Léon Brillouin, CEA-Saclay, 91191 Gif-Sur-Yvette, France

[4]Laboratoire des Matériaux et du Génie Physique, Institut National Polytechnique Grenoble, CNRS, MINATEC, 3 parvis Louis Néel, 38016 Grenoble, France

* Corresponding author: helene.bea@thalesgroup.com

A previous study of the growth conditions has shown that single-phase BiFeO$_3$ thin films can only be obtained in a narrow pressure-temperature window and that these films display a weak magnetic moment. Here we study in more detail the structure and the magnetism of single-phase BiFeO$_3$ films by means of reciprocal space mapping, Raman spectroscopy and neutron diffraction. X-ray and Raman data suggest that the BiFeO$_3$ structure is tetragonal for 70 nm-thick films and changes to monoclinic for 240 nm-thick films, thus remaining different from that of the bulk (rhombohedral) structure. In the 240 nm monoclinically distorted film neutron diffraction experiments allow the observation of a G-type antiferromagnetic order as in bulk single crystals. However, the satellite peaks associated with the long-wavelength cycloid present in bulk BiFeO$_3$ are not observed. The relevance of these findings for the exploitation of the magnetoelectric properties of BiFeO$_3$ is discussed.

## I. Introduction

In the last few years, multiferroic materials have attracted a lot of attention due to their potential applications in spintronics for instance (see introductory paper). For such applications, room temperature properties are a crucial need. Bismuth ferrite (BiFeO$_3$, BFO) is one of the few multiferroics having ordering temperatures above 300K. In bulk, it has a distorted perovskite structure, is ferroelectric up to T$_C$=1043K [1] and antiferromagnetic up to T$_N$=647K [2], with a superimposed cycloidal modulation with a period of $\lambda_{bulk}$=62 nm [3]. The main interest of BFO, besides its remarkable ferroelectric characteristics [4], resides in the possible use of its magnetoelectric properties that could allow to write a magnetic information electrically [5,6,7]. However, in bulk BFO, the linear magnetoelectric effect averages to zero due to the presence of the cycloidal order. Consequently, only the quadratic magnetoelectric effect is observed (at least at low magnetic field) [8]. Since quadratic effects are usually smaller than linear ones, especially in the range of parameters (E, H) of interest for applications, several approaches have been considered to unwind the cycloid and hence turn on the linear magnetoelectric effect. It was found that the cycloid can be destroyed by applying a magnetic field of 20T in bulk BFO [9] but also by chemical substitution at the Fe site [10]. In addition, it has been predicted that epitaxial strains can destabilize the cycloidal order and drive a transition towards a homogenous weakly ferromagnetic order [11]. Magnetization measurements on strained (111)-oriented films support this picture [11] but there is no direct evidence for the destruction of the cycloidal order, from which a linear magnetoelectric effect should emerge, in thin films.

Here, we report on X-ray diffraction, Raman spectroscopy and neutron diffraction experiments on epitaxial (001)-oriented BiFeO$_3$ thin films. We have carefully characterized the structural properties of our films and provide direct evidence that monoclinically distorted films show a G-type antiferromagnetic ordering with no indication of any cycloidal modulation. Such films are thus suitable for the exploitation of the linear magnetoelectric effect at room temperature.

**II. Structural properties**

We have grown high quality single-phased BFO films by pulsed laser deposition on (001) SrTiO$_3$ (STO) substrates using a frequency-tripled $\lambda$=355 nm Nd:YAG laser at a frequency of 2.5 Hz. We have explored different oxygen growth pressures and temperatures in order to determine the optimal growth conditions. We have observed that 70 nm BFO films are single phased only in a narrow pressure-temperature window around 6 10$^{-3}$ mbar and 580°C respectively [12]. In these growth conditions a weak magnetic moment is observed. The rms surface roughness of the films, measured by atomic force microscopy, was found to be 3.2 nm for the 70 nm film, 6 nm for the 120 nm one and 6.8 nm for the 240 nm film.

Bulk BFO is rhombohedral with pseudo-cubic lattice parameters of a = 3.96 Å and $\alpha$ = 89.4°. For low enough thickness, the structure and symmetry of epitaxial BFO films grown on cubic STO (a = 3.905 Å) substrate are thus expected to be modified compared to those of bulk BFO. X-ray diffraction allows to check the presence of parasitic phases [13], but also to study the symmetry and strain state via reciprocal space mappings (RSM). RSM around the asymmetric (103) reflection of BFO and STO were collected on films grown in the optimal conditions (fig 1). In these RSM, the vertical direction, (q$_\perp$) corresponds to 3c* and the horizontal one (q$_{//}$) to a*. For a 70 nm-film (fig 1a), the spot corresponding to BFO has the same a*, i.e. the same in-plane parameter, as the STO substrate, indicating that BFO is fully strained. $\phi$-scans were performed on this sample (fig 1d) and show 4 single peaks separated by 90° and aligned with the substrate positions (as shown in the inset), meaning that the in-plane structure is square. These data are in favour of a tetragonal unit-cell for this 70nm film. For the 240nm films, the RSM show the presence of two peaks for the BFO (fig 1c).

Furthermore, in ϕ-scans we observe four sets of double peaks (see inset) separated by 90° and aligned with the substrate. This is consistent with a monoclinic symmetry as observed by Xu et al. on 200 nm (001) BFO films [14]. If we now have a loook at the data for the intermediate thickness of 120 nm, the RSM shows a slight shift of the in-plane parameter indicating a relaxation of the film. Moreover, the peak looks distorted compared to the 70 nm one. This can be due to the presence of two domains as in the 240 nm film, but with a much lower distortion. On the other hand, the ϕ-scans for this sample show only 4 single peaks separated by 90°. This intermediate structure is thus not easily deduced from the XRD data.

### III. Raman spectroscopy

In order to better determine the symmetry changes of BFO films observed by XRD, room temperature confocal Raman spectra have been measured in a back-scattering geometry by using a LABRAM Jobin-Yvon spectrometer (naturally polarized $Ar^+$ ion laser, $\lambda = 514.53$ nm). We have checked that the laser power does not heat or modify the sample. Polarized $Z(XX)\overline{Z}$ and unpolarized Raman spectra have been collected. Below ~100 $cm^{-1}$, the decrease in intensity is due to the cut off of a Notch filter (suppression of the Raleigh scattering).

Unpolarized and polarized spectra (figure 2a and b, respectively) were measured for 70, 120 and 240 nm-thick films. For all three thicknesses the unpolarized and polarized Raman spectra are distinct, attesting a good crystalline quality. In all spectra, a large contribution due to the STO substrate is observed between 230 and 500 $cm^{-1}$ and partly masks the potential BFO bands. However, a striking observation is that both the unpolarized and polarized Raman spectra are different from those of bulk BFO ([15] and see figure 2a). This spectral modification cannot be explained by an overlap with the response from STO as illustrated by the comparison from bulk BFO and the STO substrate. In R3c bulk BFO, symmetry considerations give 13

Raman-active modes, most of them being observed in bulk spectra [15] while for our films only few bands are observed. In addition, bands not observed in the bulk are present in the spectra of our films. A first implication is that none of our films has a bulk-like rhombohedral symmetry. A second consequence is that it is difficult to give a correspondence between the observed bands and phonon modes.

However, we can try to compare qualitatively the spectra and get some insight on the BFO symmetry changes. For the unpolarized Raman spectra, small differences are visible between films of different thicknesses. Indeed, in all films, two sharp bands are visible around 140 and 175 cm$^{-1}$. For $t = 120$ and 240 nm, two small shoulders appear around 218 and 278 cm$^{-1}$ but are overlapped by the scattering from the substrate. In the $Z(XX)\overline{Z}$ polarized spectrum of the 70 nm film (fig 3b), the 140 and 175 cm$^{-1}$ bands have disappeared, while in the 240 nm polarized spectra these peaks are intense, and this cannot be attributed to a better separation of the film bands from the STO contribution. This means that the selection rule for these two peaks in the two films is not the same. This is a direct proof that the symmetry has changed between the 70 and 240 nm film. Moreover, the polarized spectrum of the 240 nm film shows a larger number of bands (at 140, 175, 220, 275, 370, 405, 470 and 540 cm$^{-1}$) compared to the 70 nm film. The larger number of peaks in the 240 nm film may reflect a space group with lower symmetry than for the thinner films, as in the case of a monoclinic symmetry compared a tetragonal one. It is then interesting to study look more carefully at the 120 nm film spectrum. In this spectrum, the two peaks at 140 and 175 cm$^{-1}$ are present, but their intensity is much lower than for the 240 nm film, contrary to the unpolarized spectra. This may reflect the presence of regions with different symmetries in the film. A first type of regions would be tetragonal and accordingly these two peaks would be absent in the polarized spectrum. The other type of regions would be monoclinic with these two peaks present. Finally, we note that Singh et al [16] have reported Raman spectra for monoclinically distorted 600 nm BFO films grown on STO (001). Their experimental results are qualitatively similar to what we have for our 240 nm film.

To summarize this structural part, our XRD and Raman data are consistent with a transition from a tetragonal symmetry for the 70 nm film to a monoclinic symmetry for the 240 nm film. The symmetry of the 120 nm films was difficult to determine with XRD experiments. In view of the Raman data, we suggest that this film consists of a mixture of tetragonal and monoclinic regions.

**IV. Magnetometry**

Magnetic hysteresis loops were measured at room temperature on a 240 nm BFO film by using an Alternating Gradient Field Magnetometer (AGFM). The measurements were made with the magnetic field in the sample plane (IP) and out of plane (OP). The raw M(H) data shown in figures 3a and 3b are dominated by the diamagnetic contribution of both the STO substrate and the sample holder, possibly masking a positive susceptibility contribution arising from the antiferromagnetic character of the BFO film. Therefore, this latter contribution, observed in bulk BFO [11], cannot be extracted from our AGFM measurements. The slopes in the IP and OP measurements are not the same because the sample holders are different in the two configurations.

In the IP measurement, a visible hysteresis is superimposed to the negative slope, while this hysteresis disappears in the OP measurement. This indicates that the BFO film has a weak ferromagnetic moment with an easy magnetization axis lying in the plane. In order to estimate the magnetization of the film, we plot in figure 3c a M(H) cycle corrected from the negative slope (calculated between 5 to 10 kOe) and obtain a value of 0.012 $\mu_B$/Fe (1.8emu/cm$^3$). This weak magnetization value is an order of magnitude higher than the value of obtained in bulk crystal at a magnetic field of 1T [17] but compares well to the extrapolated value from measurements at filed high enough to destroy the cycloid [17]. Following Bai et al [11], we may thus suppose that in our films the cycloid, that averages the magnetic moment to zero in the bulk, is broken and induces the onset of a weak magnetic moment.

**V. Neutron diffraction**

To get more insight on the microscopic magnetic order, we have performed neutron diffraction experiments on our BFO films with the triple axis 4F1 spectrometer at the LLB-Orphée reactor. Incident and final wavevectors were the same and chosen equal to $k_i = k_f = k = 1.55$ Å$^{-1}$ or 1.20 Å$^{-1}$. A cold beryllium filter was placed on the scattered beam to avoid high order contamination. The sample was placed in a vacuum can in order to minimize the background, and oriented so as to have [110]* and [001]* in the horizontal plane. Data were taken at room temperature.

Early neutron diffraction experiments on bulk BiFeO$_3$ have revealed that it is a G-type AF [18], with a superimposed cycloidal modulation [3]. The G-type order was deduced from the observation of magnetic Bragg peaks for *h-k+l=3n* with *l* odd, in hexagonal notation. In the following, we shall convert these hexagonal indices to pseudo-cubic which are simpler to handle (see table I). In those experiments, the cycloid modulation was evidenced by the presence of satellites located at [-½ -½ ½]* ± [δ,0,-δ]*, with δ=0.0045, indicating a period of about λ=62 nm [3,19] with a modulation vector $q_{bulk}$=[110]*. The intensity of these satellites turned out to be the same as the main peak.

So as to observe the antiferromagnetic order, we have focused on [½ ½ ½ ]* and [-½ -½ ½]* positions, where no structural contribution is expected. Figure 4a shows a scan carried across [½ ½ ½ ]* with k =1.55 Å$^{-1}$ and figures 4b and 4c show scans close to the [-½ -½ ½]* reflection with k =1.55 Å$^{-1}$ (Fig 4b) and k =1.20 Å$^{-1}$, for which the experimental resolution is better (Fig. 4c). All these scans show unambiguously the existence of Bragg peaks and are consistent with G-type antiferromagnetism.

Now let us discuss the possible existence of a cycloidal modulation in our films. The bold lines in figures 4b and 4c are simulations of the expected spectra, taking into account the wavevector-dependent instrumental resolution, if a cycloid with the same period as in the bulk were present. We note that for k =1.55 Å$^{-1}$ the presence of such a cycloid would considerably broaden the peak and for k =1.20 Å$^{-1}$ satellites would be clearly visible. The absence of these peaks in the experimental data of figure 4c indicates the absence of any bulk-like cycloidal modulation in our films.

Next, we would like to discuss the line width of the experimental peaks, which is a measure of the magnetic coherence length $\zeta_m$ in the samples [20]. When instrumental resolution is taken into account, we find similar values of $\zeta_m = 35.5 \pm 2.4$ nm for the three peaks of Figures 4a, 4b and 4c. Since the intrinsic width of nuclear peaks yields a coherence of $\zeta_n \approx 200$ nm (i.e. close to the film total thickness), $\zeta_m$ is not limited by structural defects but rather by the presence of magnetic domains with different orientations of the antiferromagnetic vector [21]. It is important to note that $\zeta_m < \lambda_{bulk}$, which confirms that the bulk-like cycloidal modulation is absent. However, we may suppose that a cycloid with a shorter period $\lambda_{film} \leq \zeta_m < \lambda_{bulk}$ might be present in the films. In this case the satellite peaks would be further separated from the main peak and thus even more visible in the data. The shape of the experimental spectra allows ruling out this hypothesis. Finally, a last possibility might be the presence in each domain with size $\zeta_m$ of rotating spins within some sort of incomplete cycloid. The fact that $\zeta_m$ for [½ ½ ½ ]* (i.e. perpendicular to the cycloidal modulation vector $\mathbf{q_{bulk}}$) and for [-½ -½ ½]* (i.e. a direction with a non-zero projection along the cycloidal modulation vector $\mathbf{q_{bulk}}$) are virtually the same makes this last possibility unlikely.

As previously mentioned, it has been argued that a strain of 0.5% is enough to break the cycloidal order [11]. In our 240 nm film, we estimate the strain to 1.5 %, which actually supports this picture. However, we emphasize that a definitive conclusion on the influence of strain on the magnetic order in BFO films should only be provided by systematic neutron diffraction or nuclear magnetic resonance studies of films with different strain states and magnetic coherence length (domain size). Indeed, point defects (such as Bi or O vacancies), changes in the structural symmetry or simply dimension reductions promoting a lower domain size may also be relevant.

**VI. Summary**

To conclude, we have shown by combining X-ray diffraction and Raman experiments that the apparent structural symmetry of epitaxial BFO thin films grown on STO (001) substrates changes from tetragonal to monoclinic as thickness increases. Neutron diffraction measurements reveal that 240 nm monoclinically distorted BFO films are G-type antiferromagnets but that the incommensurate cycloid present in the bulk is absent, presumably due to epitaxial strain. We emphasize that these films therefore fulfil the requisite for the observation of a linear magnetoelectric effect, prohibited in the bulk. This condition, together with the observation of a monoclinic space group, appears to be a key ingredient for a clean control of the antiferromagnetic vector by electric field in BFO structures [6]. The next steps towards achieving a better understanding of this magnetoelectric effect and of the recently observed exchange-bias [7,22] are to unveil the relation between film thickness and the type of antiferromagnetic order, and the precise orientation of the antiferromagnetic vector within the domains. To reach these goals, neutron diffraction appears as a very powerful technique, more direct than X-ray linear magnetic dichroism as it can probe magnetic order only.

In the process of revising this article, Saito *et al* [23] reported a transition from tetragonal symmetry to monoclinic symmetry as thickness increases above about 50 nm in (001)BFO films, as inferred from RSM data. Our XRD and Raman spectroscopy data bring additional evidence of this effect and confirms that even very thick (001-)-oriented BFO films do not recover a bulk-like rhombohedral symmetry.

This work has been supported by the E.U. STREP MACOMUFI (033221), the contract FEMMES of the Agence Nationale pour la Recherche and the GDR FIFA. H.B. also acknowledges financial support from the Conseil Général de l'Essonne. The authors would also like to thank A.K. Zvezdin and J.F. Scott for fruitful discussions.

Figure Captions :

Fig 1. Reciprocal Space Mapping around the (103) reflection of (a) 70 nm, (b) 120 nm and (c) 240 nm BFO films. Phi-scans of the (202) reflection of BFO for 70 nm (d), 120 nm (e) and 240 nm (f) films. Insets show in more detail one of the peaks of BFO and STO with pseudo-Voigt fits (red thick lines) with a single line (d and e) or two lines (f) for the BFO peak.

Fig 2. Normalized unpolarized (a) and polarized (b) Raman spectra for 70 nm, 120 nm, 240 nm BFO films, STO substrate measured at room temperature. For comparison, we add the unpolarized spectra of bulk BFO (from [15]).

Fig 3. Raw magnetization versus field measured with an AGFM at room temperature for a 240 nm film with the field in plane (a) and out of plane (b). (c) Magnetization versus field corrected with the diamagnetic slope for the in plane field.

Fig 4. Neutrons diffraction patterns of $[½ ½ ½ ]^*$ (a) and $[-½ -½ ½]^*$ (b) and (c) measured along the $[1\ 1\ 1]^*$ and $[-1\ -1\ 1]$ directions respectively for a 240 nm BFO film. The input and output collimation values are 60'. The wavectors used are 1.55 Å$^{-1}$ for (a) and (b) and 1.2 Å$^{-1}$ for (c). The dotted lines are simulated main and satellite peaks. The sum of their intensities is shown as a bold blue line. The red line is a gaussian fit to the data.

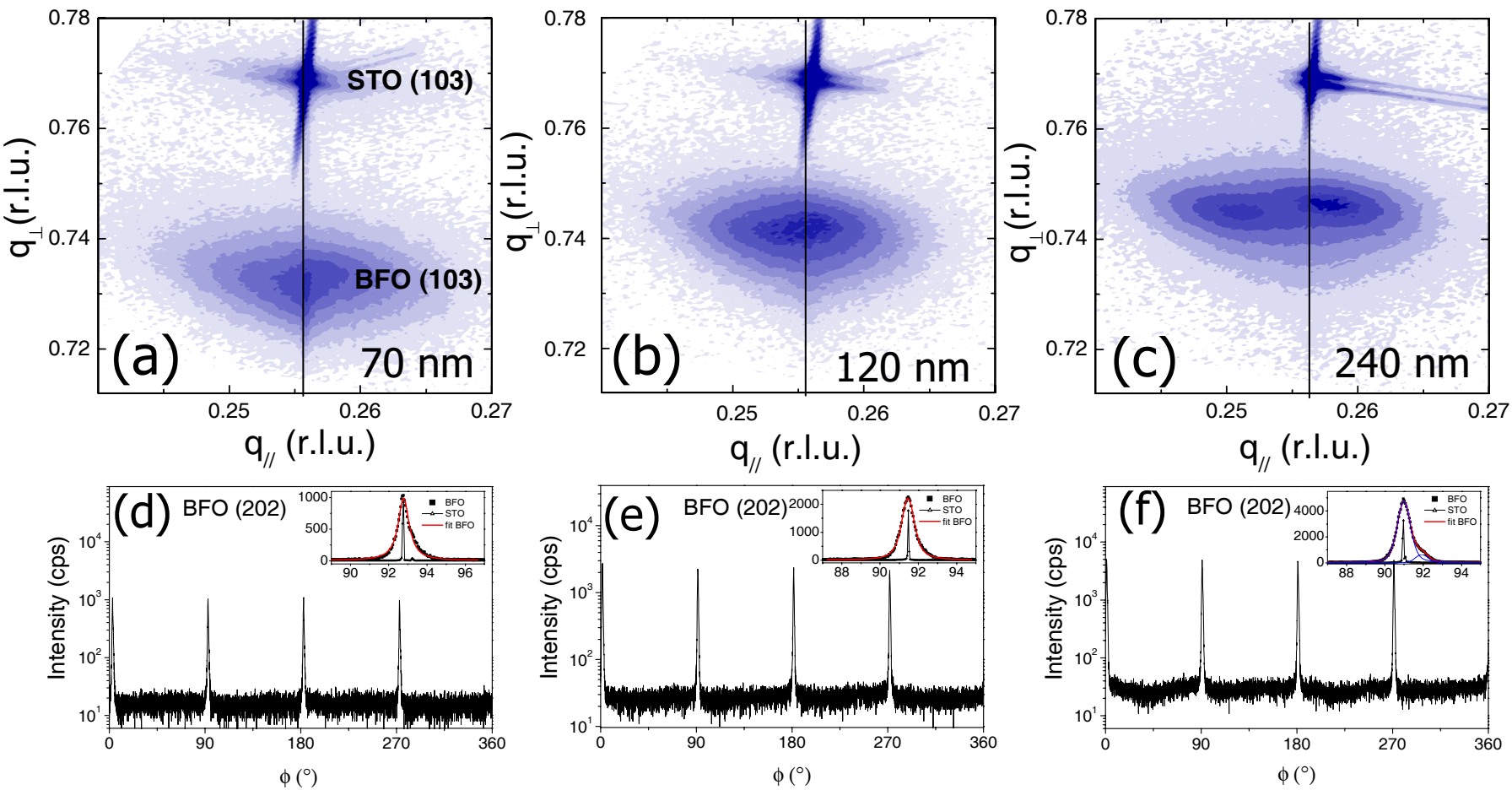

H. Béa et al; Fig.1

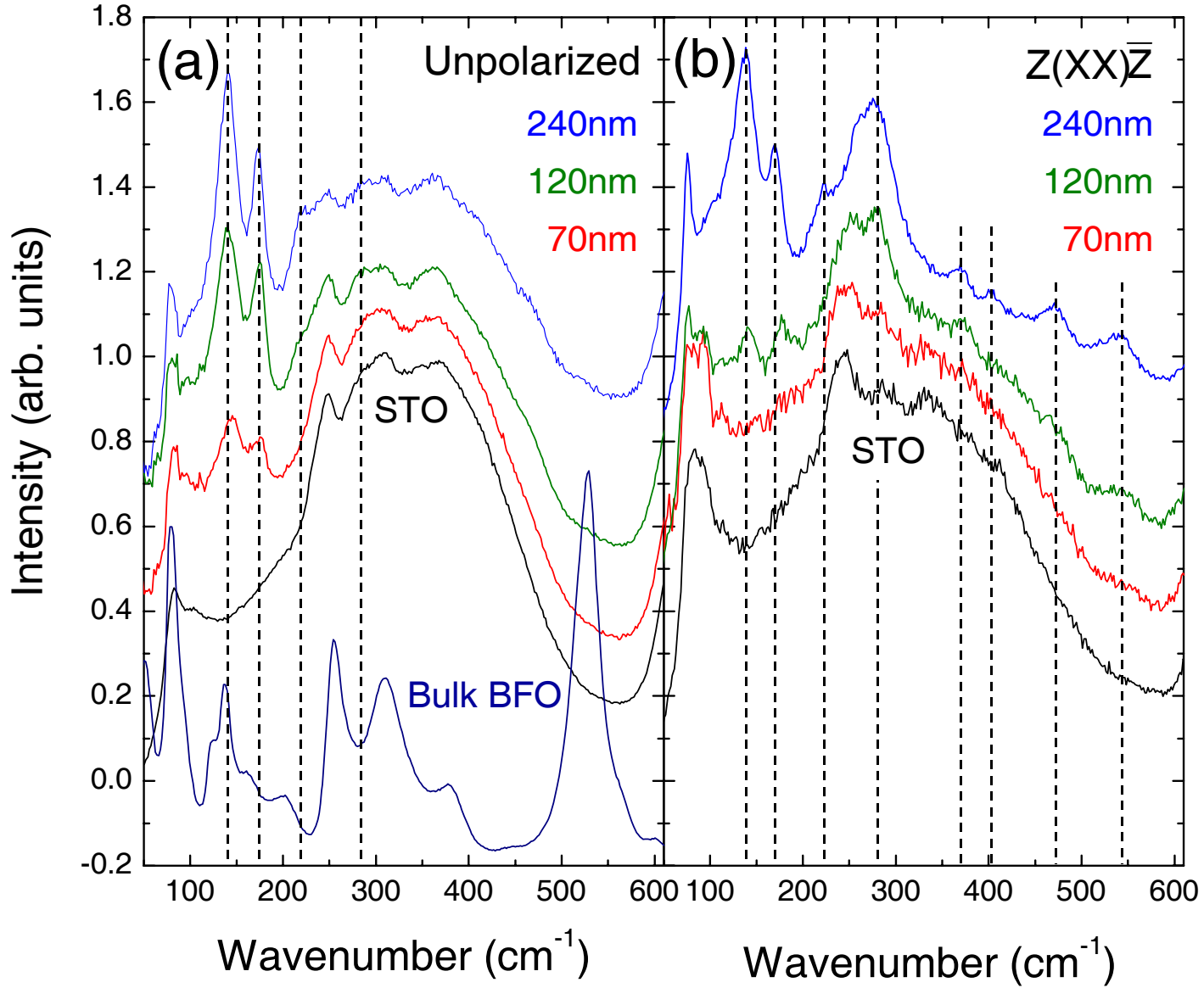

H. Béa et al; Fig.2

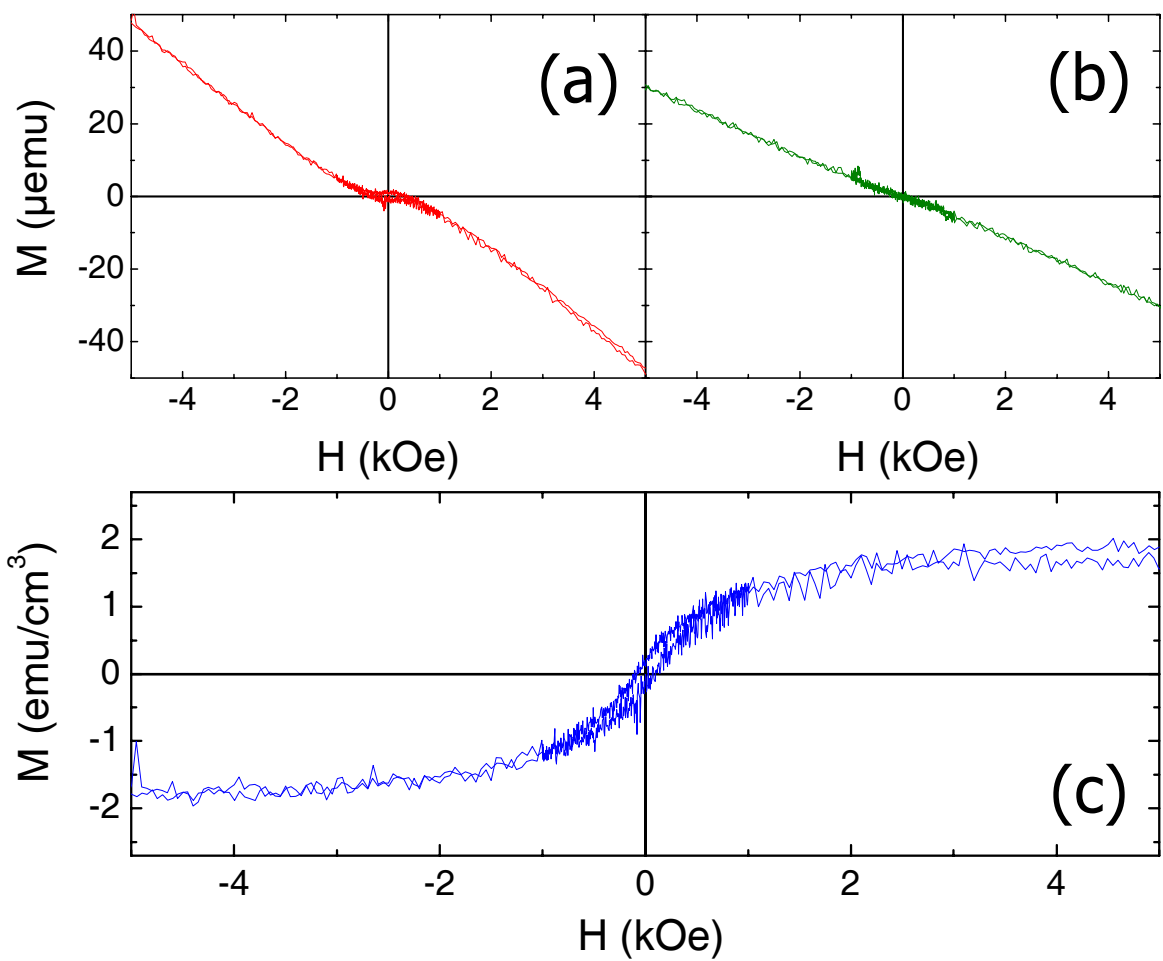

H. Béa et al; Fig.3

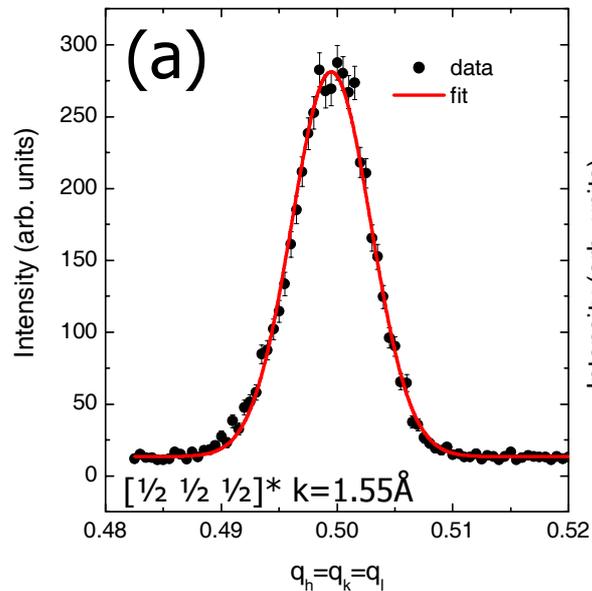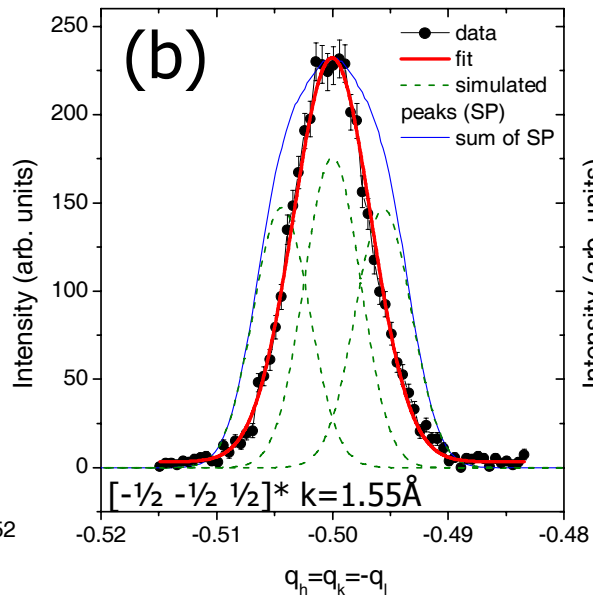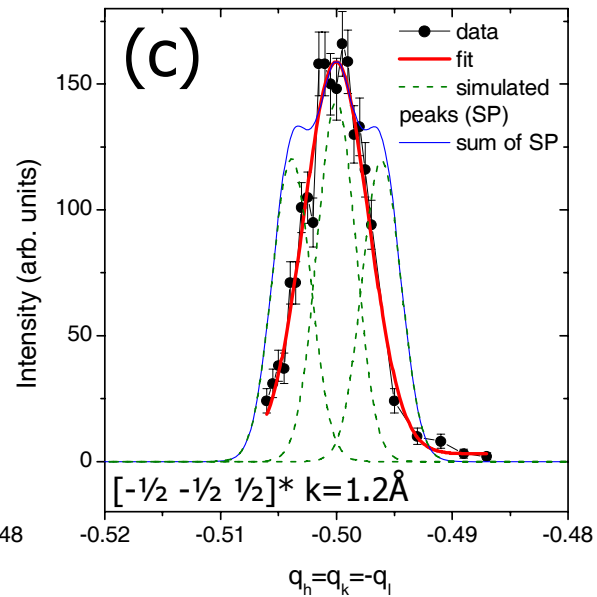

H. Béa et al; Fig.4